\documentclass[twocolumn]{aastex63}

\def\lapp{\ifmmode\stackrel{<}{_{\sim}}\else$\stackrel{<}{_{\sim}}$\fi}
\def\gapp{\ifmmode\stackrel{>}{_{\sim}}\else$\stackrel{>}{_{\sim}}$\fi}
\usepackage{multirow}
\usepackage{color}
\usepackage{amsmath}
\usepackage{soul}
\usepackage{hyperref}

\shorttitle{}
\shortauthors{}
\begin{document}
\title{Gamma-ray orbital modulation of the transitioning millisecond pulsar binary XSS~J12270$-$4859}

\correspondingauthor{Hongjun An}
\email{hjan@cbnu.ac.kr}

\author{Hongjun An}
\affiliation{Department of Astronomy and Space Science,
Chungbuk National University, Cheongju, 28644, Republic of Korea}


\begin{abstract}
We report on gamma-ray orbital modulation of the transitioning millisecond pulsar binary XSS~J12270$-$4859
detected in the {\it Fermi} Large Area Telescope (LAT) data.
We use long-term optical data taken with the {\it XMM-Newton}
Optical Monitor and the {\it Swift} UltraViolet Optical Telescope
to inspect radio timing solutions that are
limited to relatively short time intervals, and find that extrapolation of the
solutions aligns well the phasing of the optical data over 15\,years.
The {\it Fermi}-LAT data folded on the timing solutions exhibit
significant modulation ($p=5\times10^{-6}$) with
a gamma-ray minimum at the inferior conjunction of the pulsar.
Intriguingly, the source seems to show similar modulation
in both the low-mass X-ray binary (LMXB) and the millisecond pulsar (MSP) states,
implying that mechanisms for gamma-ray emission in the two states are similar.
We discuss these findings and their implications using an intra-binary shock scenario.
\end{abstract}


\bigskip
\section{Introduction} \label{sec:intro}
Spider pulsar binaries (pulsar binaries hereafter)
are composed of a millisecond pulsar (MSP) and a low-mass companion,
where the companion surface is heated and evaporated by the pulsar's irradiation.
In a pulsar evolutionary scenario \citep[e.g.,][]{bv1991}, pulsar-binary systems are believed
to be descendants of low-mass X-ray binaries (LMXBs) in which the low-mass companion
loses its mass and the neutron star spins up by accretion
over a long time \citep[$\sim$Gyr;][]{acrs82,bv1991}.
When the accretion is quenched, the systems are observed as pulsar binaries.
This evolutionary scenario is supported by discoveries of objects that
have exhibited transitions between an accretion-powered LMXB
and a rotation-powered MSP state \citep[][]{asrk+09,pfbr+13,bphk+14}.

Emission mechanisms of transitioning systems in the LMXB and the MSP states differ.
In the MSP state, like in pulsar binaries, the companion is heated by the pulsar's energy output
and so exhibits day-night cycles. The heating can make the companion wind strong, and it
interacts with the pulsar wind to form an intra-binary shock \citep[IBS; e.g.,][]{whvb+17,kra19,mwvh+20}.
Pulsar wind particles are accelerated in the shock and emit synchrotron radiation, producing
characteristic double peaks in the X-ray light curve around a maximum \citep[e.g.,][]{rmgr+15}.
Gamma rays are produced primarily in the pulsar magnetosphere, but
inverse-Compton scattering (ICS) of the stellar blackbody radiation by energetic particles
in the shock or in the pre-shock pulsar wind can also produce gamma-ray radiation
in the GeV band; this ICS emission may exhibit orbital modulation due to variation
of the viewing geometry, having a maximum or minimum at the pulsar's conjunctions \citep[e.g.,][]{ark20,cnva+21}.

In the LMXB state, the optical and X-ray emissions are mostly
powered by accretion. The surface of the pulsar can be heated by channeled accretion
and so X-ray pulsations are detected \citep[][]{abph+15,pdbb+15}. Because the accretion can be unstable,
orbital light curves in the optical and X-ray bands exhibit dips and flares \citep[e.g.,][]{dbfp+13,tyak+14}.
The gamma-ray emission mechanism, other than the magnetospheric one, in this state is rather unclear. A model suggests that
the IBS is destroyed by the strong accretion and gamma rays are produced by the synchrotron-self-Compton
process at the interface between the magnetosphere and accretion disk \citep[][]{ptl14}.
Alternatively, it was suggested that ICS of disk photons by the pre-shock pulsar wind
is responsible for the gamma rays \citep[e.g.,][]{tllk+14}. In these scenarios, orbital modulation in the
gamma-ray band seems unlikely since conditions for the emission do not vary over the orbit.

XSS~J12270$-$4859 (4FGL~J1228.0$-$4853; J1227 hereafter) is
a transitioning MSP (tMSP) which transitioned from an LMXB
to an MSP state late 2012 \citep[$\sim$MJD~56250;][]{bphk+14}.
Radio timing studies \citep[][]{rrbs+15,dpbp+20}
revealed that the source has a 1.69\,ms pulsar in a 6.9\,hr orbit with a
G-type companion \citep[][]{dpbb+15}. After the transition, the source was well studied at
X-rays, and the characteristic double-peak X-ray light curve was clearly detected
with a maximum at the pulsar's inferior conjunction \citep[][]{dpbp+20},
as expected in IBS scenarios if the stellar
wind is stronger than the pulsar's \citep[e.g.,][]{kra19}.
In the gamma-ray band, the pulsar's pulsations were detected over a relatively
short time period after the transition \citep[][]{jrrc+15}, and weak orbital modulation ($\le$3$\sigma$)
in 1.6\,years of $>300$\,MeV data taken with {\it Fermi} Large
Area Telescope \citep[LAT;][]{fermimission} was suggested \citep[][; XW15 hereafter]{J1227xw15}.
In that work, the gamma-ray orbital modulation had a `maximum'
at the pulsar's inferior conjunction (IFC; pulsar between the companion and observer).
Since {\it Fermi} LAT has been collecting data for more than 12\,years, the orbital modulation may be detected
with higher confidence and characterized better. 

In this paper, we investigate the gamma-ray modulation in J1227 using 12.5\,years of
data taken with {\it Fermi} LAT.
Since radio timing solutions that we found in the literature do not cover the whole time period of the LAT data, we inspect
the timing solutions using optical data taken with {\it XMM-Newton} Optical Monitor (OM) and
{\it Swift} UltraViolet Optical Telescope (UVOT).
We use the timing solutions that adequately explain the phasing of the long-term optical data for a LAT timing analysis.
We present optical data analysis and verify the timing solutions using
the phasing of the optical light curves in \S~\ref{sec:sec2}. We report analysis results of the {\it Fermi} LAT data
in \S~\ref{sec:sec3}, and discuss in \S~\ref{sec:sec4}.

\begin{table}
\begin{center}
\caption{Optical data used in this work}
\label{ta:ta1}
\scriptsize{
\begin{tabular}{ccccc}
\hline
Inst. & Filter & Nobs & Obs. ID & Time (MJD)\\  \hline
UVOT &   v    &  18  & 35101001--82011001 & 53628--57195 \\
UVOT &   u    &  29  & 35101003--95612003 & 56319--59184 \\
OM   &   U    &  1   & 551430401 & 54837 \\
OM   &   U    &  1   & 656780901 & 55562 \\ 
OM  &   U    &  1   & 727961401 & 56656 \\
OM   &   U    &  1   & 729560801 & 56835 \\ \hline
\end{tabular}}\\
\end{center}
\end{table}

\section{Optical Data Analyses} \label{sec:sec2}
Orbital periods of pulsar binaries can vary slightly over time, and
radio timing solutions for J1227
obtained over short time intervals (c.f., $>$12.5\,year of LAT operation)
require one or two orbital-period derivatives \citep[][]{rrbs+15,dpbp+20}.
They may reflect long-term changes of the orbital motion
or a small but unpredictable variation as seen in other pulsar binaries
\citep[e.g., $\Delta P_{\rm B}/P_{\rm B}\le 10^{-6}$ for PSR~J2039$-$5617;][]{cnva+21}.
In the latter case, the short-term radio timing solutions may be valid only within the given intervals.
Hence we check to see if the
`orbital' (not the pulsar spin) solutions are valid over the 12.5-yr period.
This can be done by carefully phasing the
day-night cycles of the companion emission which is expected to be maximum
at the binary phase $\phi_{\rm B}\approx 0.75$ (defined to be IFC).
For this test, we use optical photometric data collected with
the {\it XMM-Newton} OM and the {\it Swift} UVOT over a period of 15\,years (MJD~53628--59184).

\begin{figure*}
\centering
\begin{tabular}{ccc}
\includegraphics[width=2.2 in]{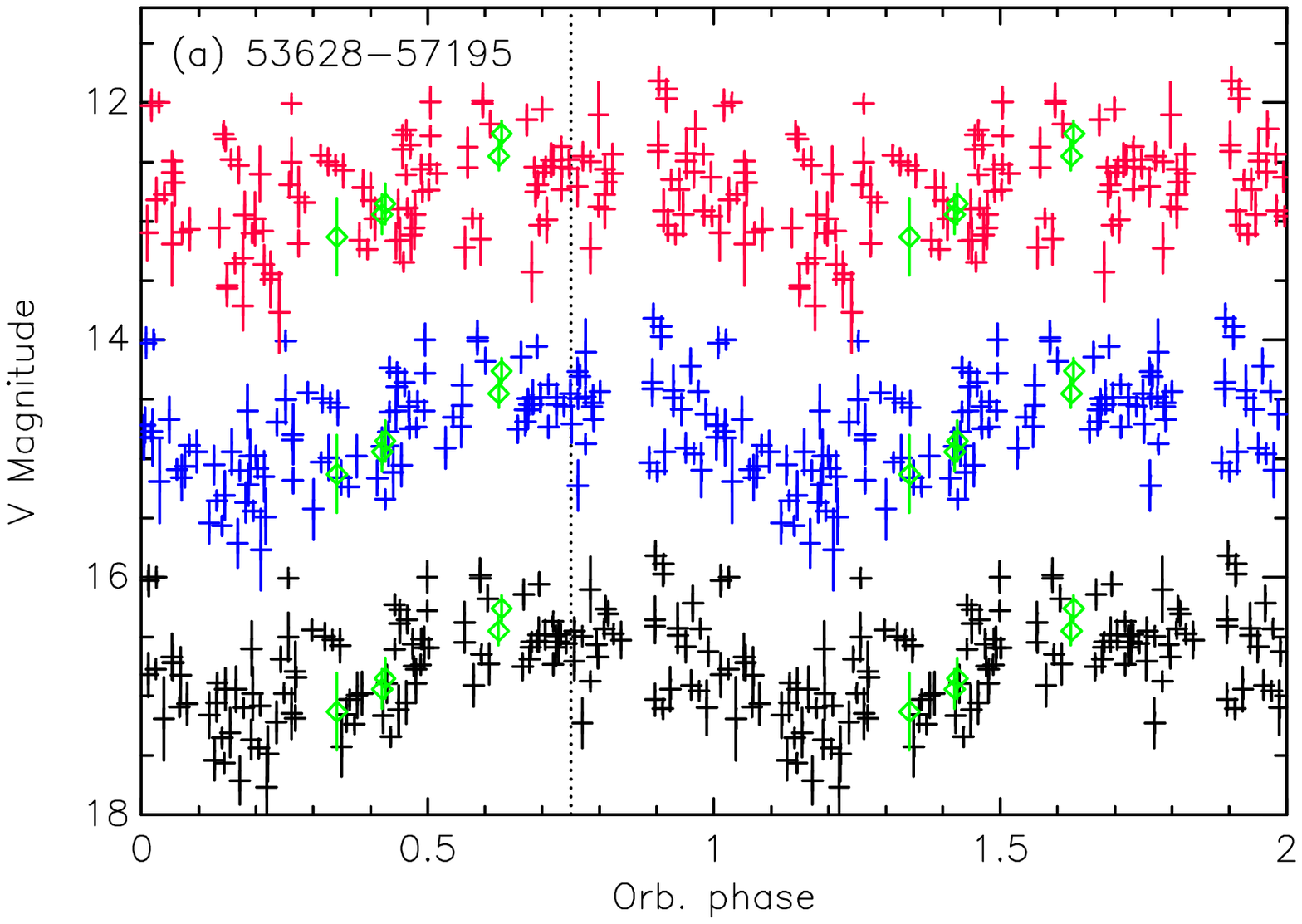} &
\includegraphics[width=2.2 in]{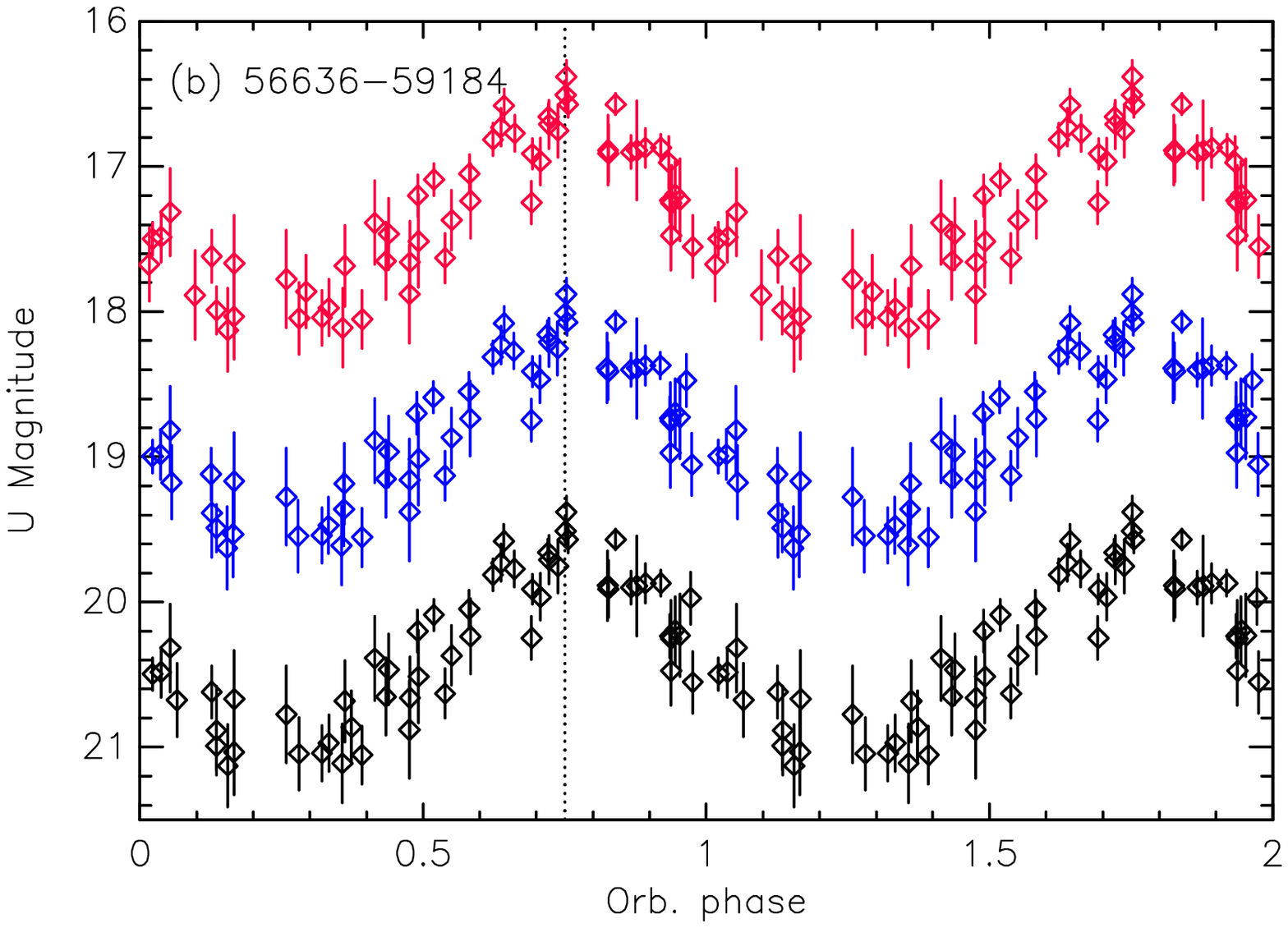} &
\includegraphics[width=2.2 in]{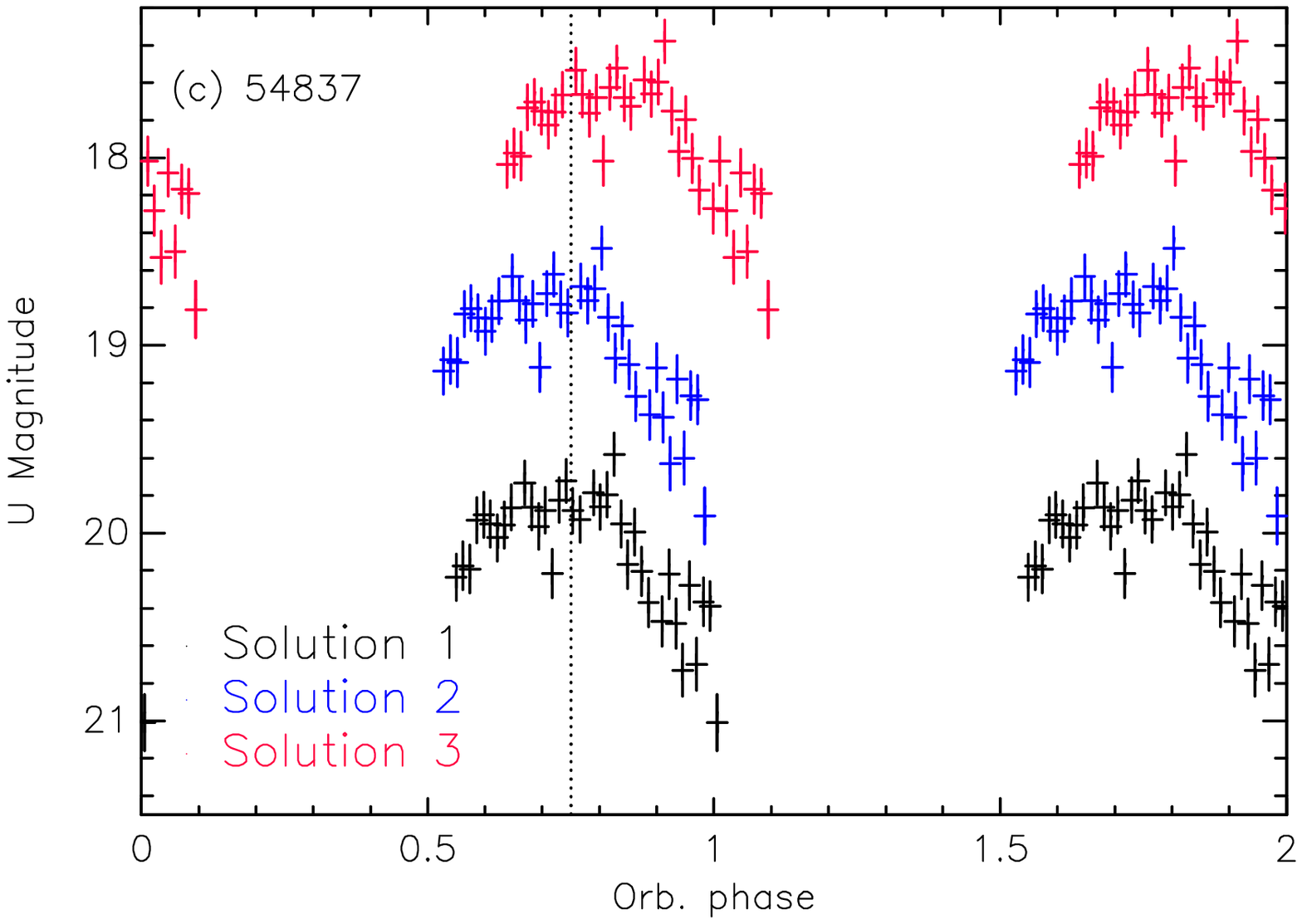} \\
\includegraphics[width=2.2 in]{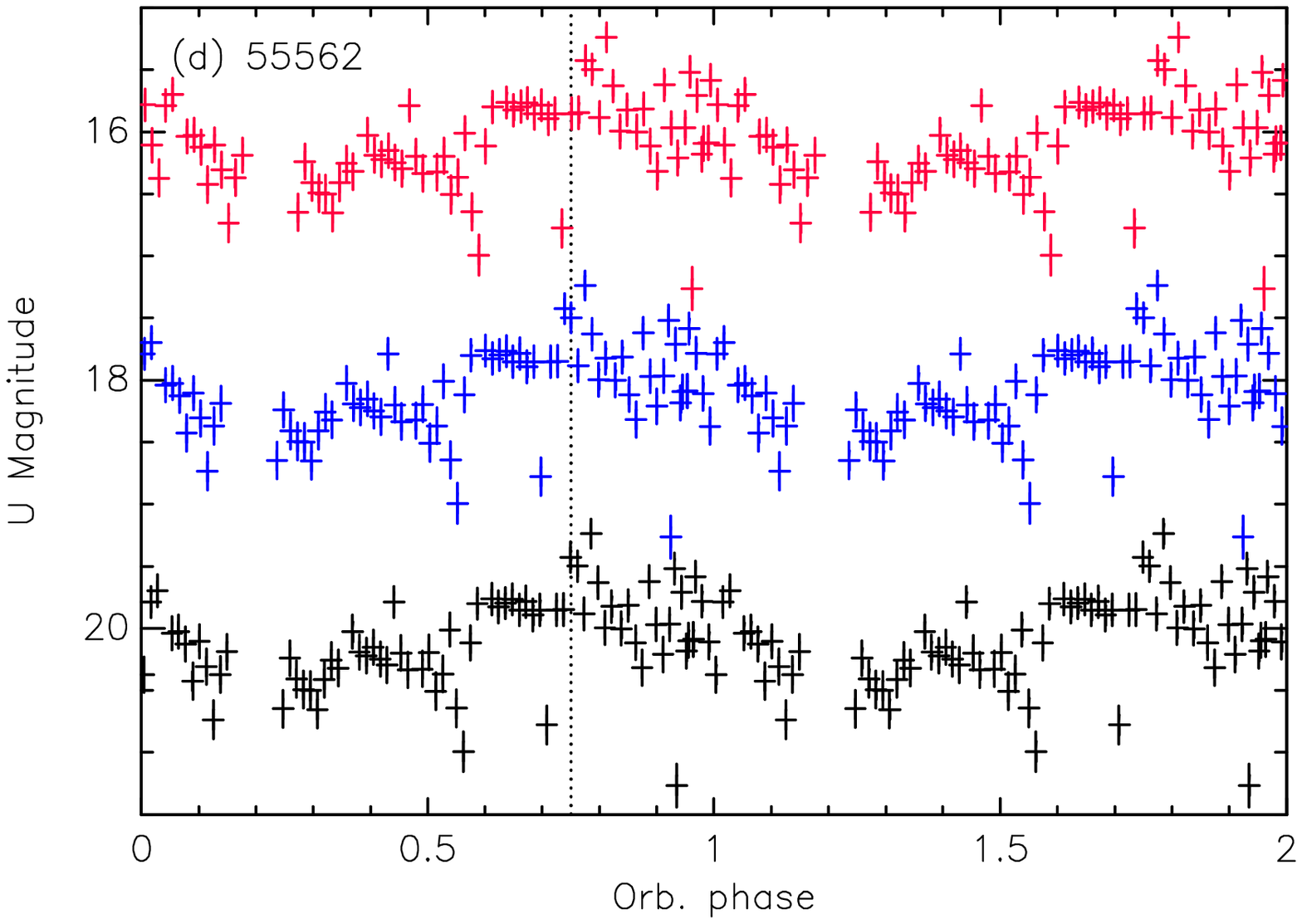} &
\includegraphics[width=2.2 in]{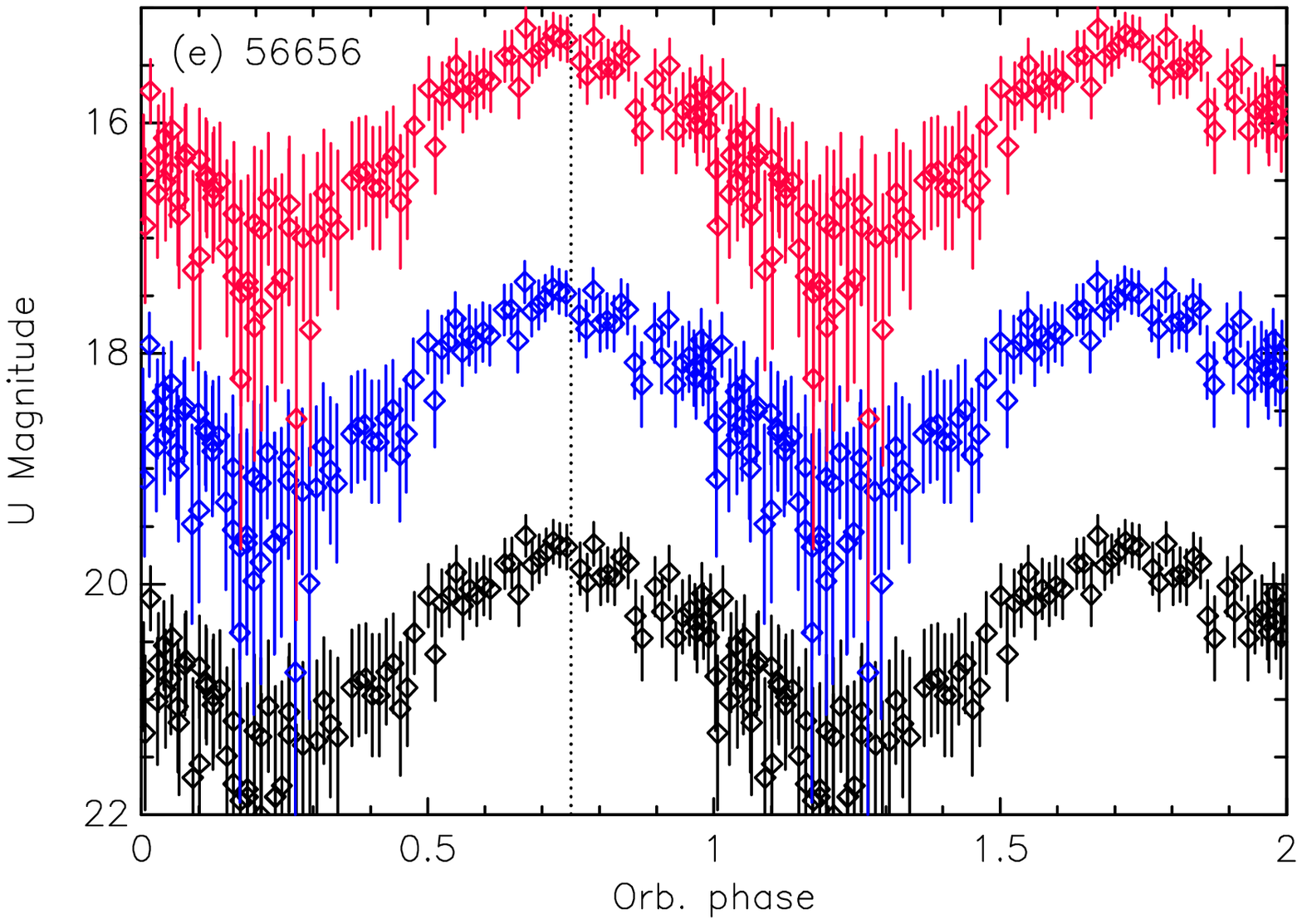} &
\includegraphics[width=2.2 in]{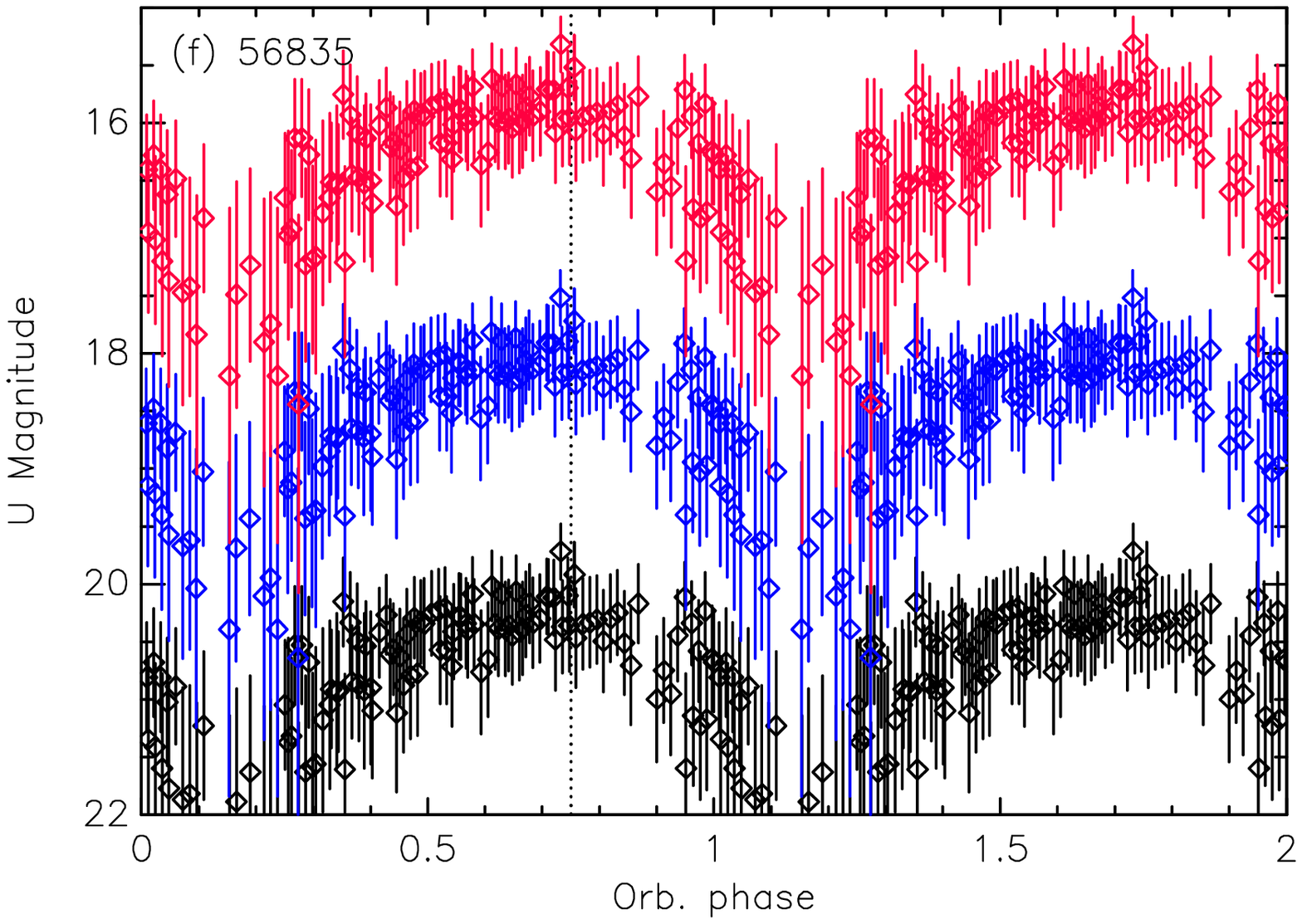} \\
\end{tabular}
\figcaption{Optical light curves folded on timing solutions found in literature
(black, blue, and red for Solutions~1, 2 and 3, respectively; Table~\ref{ta:ta2}).
The blue and red light curves are displaced vertically, and
pre- and post-transition data are plotted in crosses and diamonds, respectively.
Phase 0 is the ascending node, and the vertical dotted line denotes the IFC phase.
({\it a}): {\it Swift} v-band data. Most of the v-band data are taken before the transition ($<$MJD~56250).
Five data points plotted in green are taken after the transition, have higher magnitudes than the
pre-transition ones, and are vertically displaced.
({\it b}): {\it Swift} u-band data.
({\it c--f}): {\it XMM-Newton} U-band data on MJD~54837 (c), MJD~55562 (d), MJD~56656 (e) and MJD~56835 (f).
\label{fig:fig1}
}
\vspace{0mm}
\end{figure*}

\subsection{Optical Data Reduction} \label{sec:sec2_1}
We use {\it Swift} and {\it XMM-Newton} data taken between 2005 Sep. and 2020 Dec.
There are 63 {\it Swift} and 4 {\it XMM-Newton} observations in the period (Table~\ref{ta:ta1}).
The {\it Swift} UVOT observations are made with various filters, but we use the v- or u-filter data only
since the source is well detected in those bands. The {\it XMM-Newton} data are all made
with the U filter.

For the {\it Swift} data analysis, we use an $R=5''$ aperture for the source and an $R=7.5''$ aperture
in a source-free region for the background extraction. We process the data with {\tt HEASoft}~v6.27
following the standard UVOT analysis procedure\footnote{https://www.swift.ac.uk/analysis/uvot/}
to measure the AB magnitude in each frame.
The observation times are barycenter-corrected using
the barycenter-corrected XRT arrival times in each observation for the source position R.A.=$186.99468^\circ$
and Decl.=$-48.89521^\circ$ \citep[][]{rrbs+15}.

We process the {\it XMM-Newton} OM data with XMMSAS~v20190531\_1155 following the standard procedure.
\footnote{https://www.cosmos.esa.int/web/xmm-newton/sas-thread-omf} The arrival times are barycenter-corrected
with the {\tt barycen} tool of SAS, and the measured count rates
are converted into AB magnitudes.\footnote{https://www.cosmos.esa.int/web/xmm-newton/sas-watchout-uvflux}
The light curves are then constructed with 300-s time bins. 

Figure~\ref{fig:fig1} shows {\it Swift} and {\it XMM-Newton} optical light curves folded on
the timing solutions in Table~\ref{ta:ta2}.
The light curves show well the day-night cycles of the companion emission.
The {\it Swift} data cover various epochs over 15\,years; the v- and u- band
data were taken mostly in the pre-transition and post-transition state, respectively.
Notice that pre-transition (LMXB state) light curves (panels a, c, and d) exhibit some
dips and flares.

\subsection{Inspection of Radio Timing Solutions} \label{sec:sec2_2}

\begin{table*}
\begin{center}
\caption{Radio timing solutions for the orbit of J1227 used in this work}
\label{ta:ta2}
\scriptsize{
\begin{tabular}{lcccc}
\hline
Properties    & Solution~1 & Solution~2 & Solution~3   \\  \hline
$F_{\rm B}$ (s$^{-1}$) & $4.02034590\times10^{-5}$ & $4.02035224\times10^{-5}$ & $4.02034315\times10^{-5}$ \\
$\dot F_{\rm B}$ (s$^{-2}$) & $1.406\times10^{-18}$ & $4.04\times10^{-19}$ & $3.74\times10^{-18}$ \\
$\ddot F_{\rm B}$ (s$^{-3}$) & $0$  &  $0$ & $-7.77\times10^{-26}$ \\
$T_{\rm ASC}$ (MJD)      & 56700.9070772  & 57139.0715595  &  56700.907021 \\
TSTART (MJD)  & 56707.95  & 57063.530 & 56824.26 \\
TSTOP (MJD)   & 56978.13  & 57233.225 & 57685.21 \\
$\chi^2$/dof  & 219.9/236  & 217.7/236 & 223.0/236 \\
Refs.         & \citet{rrbs+15}  & \citet{dpbp+20}  & \citet{dpbp+20} \\
\hline
\end{tabular}}\\
\end{center}
\end{table*}

\citet{rrbs+15} and \citet{dpbp+20} provided
radio timing solutions for J1227 that were derived
for relatively short intervals (Table~\ref{ta:ta2}).
From an initial inspection of the solutions for 12.5\,years of LAT operation,
we find that Solutions~1 and 2 are very similar; phase misalignment between them is only $\le 0.03$.
Solution~3 differs from the others with a phase misalignment of $\sim 0.1$ and so will be
discernible by the 15-yr optical data. Note, however, that ignoring the second derivative
from Solution~3 makes the misalignment small, at a $\sim0.05$ level.

We extrapolate the solutions to the 15-year span of the optical data.
Because the orbital frequency ($F_{\rm B}$) is not expected to vary much and
the timing solutions do not use high orbital frequency derivatives, the solutions seem to align well
with the optical phases at various epochs in the 15\,year period (Fig.~\ref{fig:fig1}), meaning that
extrapolating the radio solutions is adequate for the orbit.
However, we note that Solution~3 (red in Fig.~\ref{fig:fig1}) misaligns
some early observations; day-night cycles seen in the light curves folded on Solutions~1 and 2
are blurred (Fig.~\ref{fig:fig1} a) and the peak phase is shifted (Fig.~\ref{fig:fig1} c).
This implies that Solution~3 may not be accurate outside the validity window
because of the second derivative $\ddot F_{\rm B}$; simply ignoring it in the solution
recovers the phase alignment.

We make more quantitative checks on the timing solutions by modeling the light curves with a sine function
having time-varying frequency
\begin{equation}
M(t)=M_0\mathrm{sin}[\omega_{\rm B}(t-T_0) +\dot \omega_{\rm B}(t-T_0)^2/2 +\ddot \omega_{\rm B}(t-T_0)^3/6]+C,
\end{equation}
where $M(t)$ is the magnitude, $\omega_{\rm B}$ is $2\pi F_{\rm B}$,  and $T_0$ is the epoch of the ascending node.
Note that we do not use the pre-transition data in the fit because of the large scatter in the data caused by dips and
flares (e.g., Fig.~\ref{fig:fig1} d); small inaccuracy in the phases of these data
will bias the solution significantly due to the long baseline.
Besides, a single sine function does not fit the data across the transition because the source was substantially brighter
and the modulation amplitude appeared to be smaller in the pre-transition data than in the post-transition ones.

We fit the post-transition light curves
(Fig.~\ref{fig:fig1} b, e, and f) holding the frequency and its derivatives fixed at the values in Table~\ref{ta:ta2},
and compare $\chi^2$ of the fits for different solutions.
The $\chi^2$ values are presented in Table~\ref{ta:ta2}.
All the solutions are acceptable, but Solution~3 is inferior to the other ones.
We also vary $F_{\rm B}$ and/or $\dot F_{\rm B}$, and find that
optimizing these values does not significantly improve the fits with $\chi^2$/dof=216.3/235
($F_{\rm B}=4.02033(2)\times10^{-5}\rm \ s^{-1}$) and 215.7/234
($F_{\rm B}=4.02032(2)\times10^{-5}\rm \ s^{-1}$ and $\dot F_{\rm B}=2(2)\times 10^{-18}\rm \ s^{-2}$);
the best-fit $F_{\rm B}$ and $\dot F_{\rm B}$ are consistent with the radio solutions.
The optical data are well fit with a model having $F_{\rm B}$ only, and including higher
frequency derivatives does not provide a statistically better fit.
Note that using the optimized values of $F_{\rm B}$ and/or $\dot F_{\rm B}$ for the LAT data
does not significantly alter the results below.

\section{Fermi-LAT Data  Analyses} \label{sec:sec3}
\begin{figure}
\centering
\includegraphics[width=3.4 in]{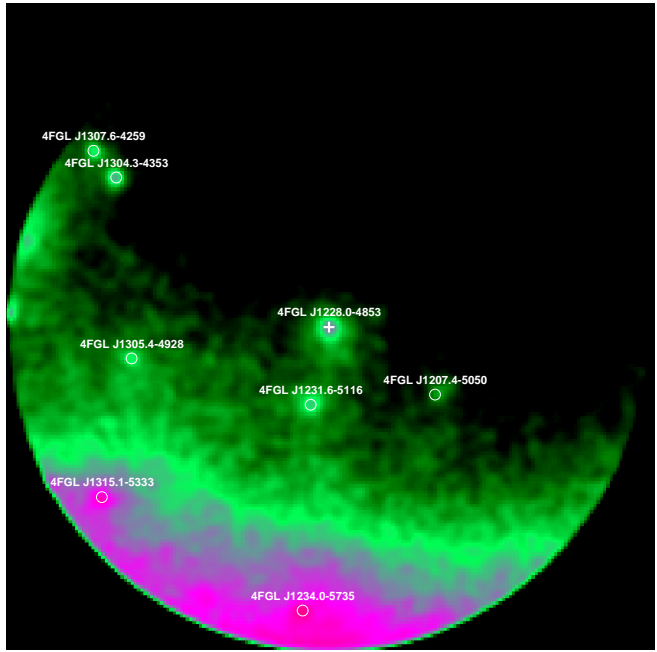} 
\figcaption{A {\it Fermi}-LAT image of a $R=10^\circ$ field in the 100\,MeV--300\,GeV band.
J1227 is marked with a cross at the center, and other bright sources are denoted in a circle.
The image is smoothed and the scale is adjusted for better legibility.
\label{fig:fig1x}
}
\vspace{0mm}
\end{figure}

We analyze {\it Fermi}-LAT data collected between 2008 Aug. and 2021 Feb. spanning approximately 12.5\,years.
The data are analyzed with the {\it Fermi} Science Tools (ST) v1.2.23 along with the {\tt P8R3\_SOURCE\_V2}
instrument response.\footnote{https://fermi.gsfc.nasa.gov/ssc}
We select the {\tt Front+Back} event type in the {\tt SOURCE} class
within an $R=10^\circ$ region of interest (RoI), and
reduce the data using the zenith angle $<90^\circ$, DATA\_QUAL$>$0, and LAT\_CONFIG=1.
A 100\,MeV--300\,GeV image is displayed in Figure~\ref{fig:fig1x}, and we further analyze the data as described below.

\subsection{Spectral Analysis} \label{sec:sec3_1}
We first perform a binned likelihood analysis in the 100\,MeV--300\,GeV band to measure the source spectrum.
Because our data are not very different from those used for the 4FGL catalog \citep[{\tt gll\_psc\_v21.fit};][]{fermi4fgl}
and the source state has not changed recently, the 4FGL values are expected to be very accurate.
Nevertheless we verify the parameter values below.

We use the same spectral models as in 4FGL and fit parameters for bright sources
within $3^\circ$ from J1227 and
amplitudes of diffuse emission\footnote{https://fermi.gsfc.nasa.gov/ssc/data/access/lat/Background\\Models.html}
({\tt gll\_iem\_v07} and {\tt iso\_P8R3\_SOURCE\_V2\_v1}).
J1227 emission is modeled with PLEXP2, $dN/dE=N_0(E/E_0)^{-\Gamma_{\rm 1}}e^{-aE^{\Gamma_{\rm 2}}}$
with $E_0$ and $\Gamma_2$ held fixed at 1.5\,GeV and 0.67, respectively.
We gradually include the next bright sources in the fit and compare the Akaike Information Criteria \citep[][]{aic74}
fit statistics. We repeat this until the fit does not improve significantly, which requires to fit five sources.
We find that the best-fit parameters for J1227 are
$N_0=(3.7\pm0.6)\times 10^{-12}\rm \ phs\ cm^{-2}\ s^{-1}\ MeV^{-1}$, $\Gamma_{\rm 1}=1.6\pm0.1$ and $a=(7\pm1)\times10^{-3}$.
Since these values are consistent with those in 4FGL at the $\le$2$\sigma$ levels,
we use the 4FGL model for the timing analysis.
We verified that using our optimized or the 4FGL-DR2 ({\tt gll\_psc\_v27.fit}) model
does not significantly alter the results below.

\begin{figure*}
\centering
\begin{tabular}{cc}
\includegraphics[width=3.3 in]{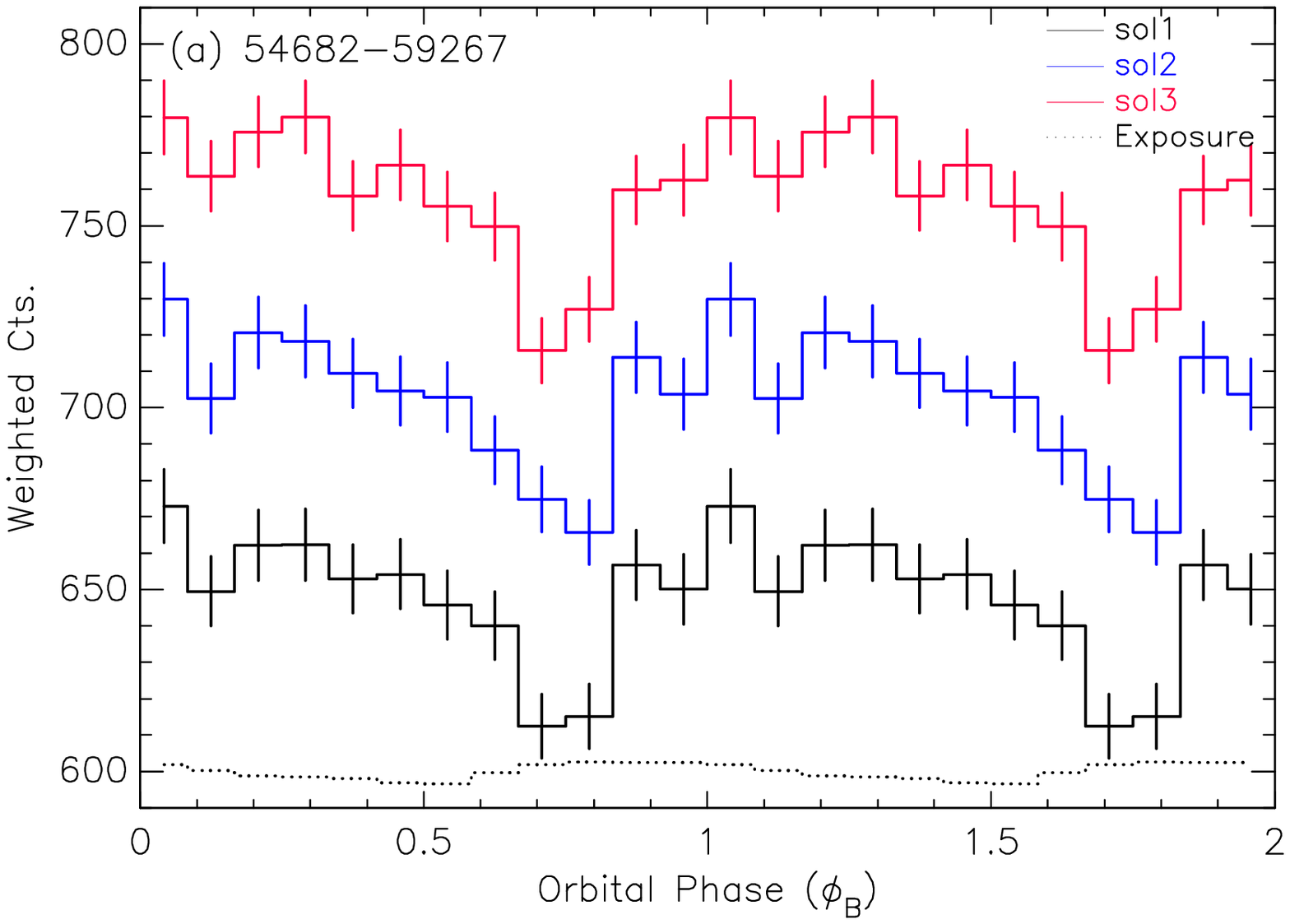} &
\includegraphics[width=3.3 in]{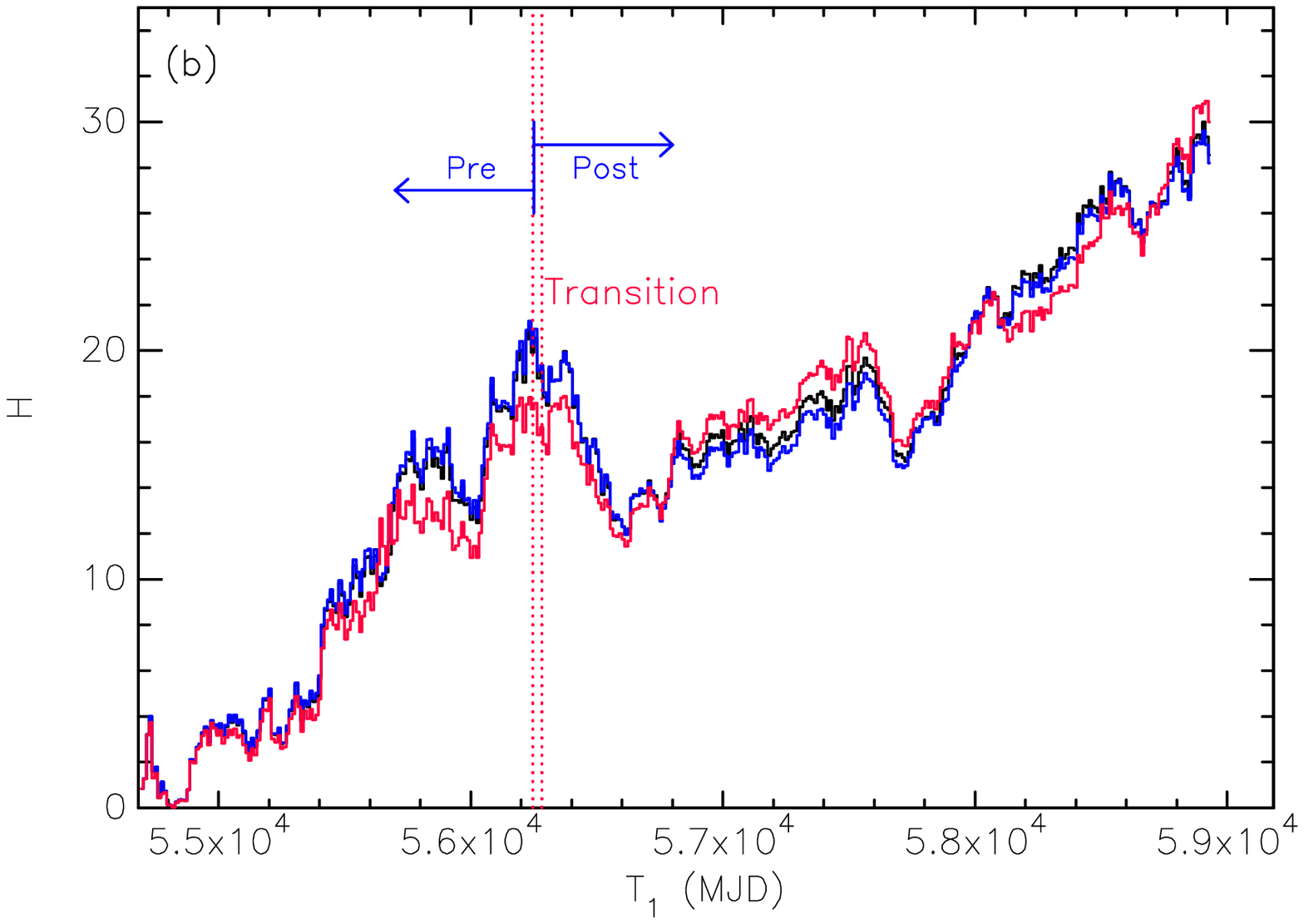} \\
\includegraphics[width=3.3 in]{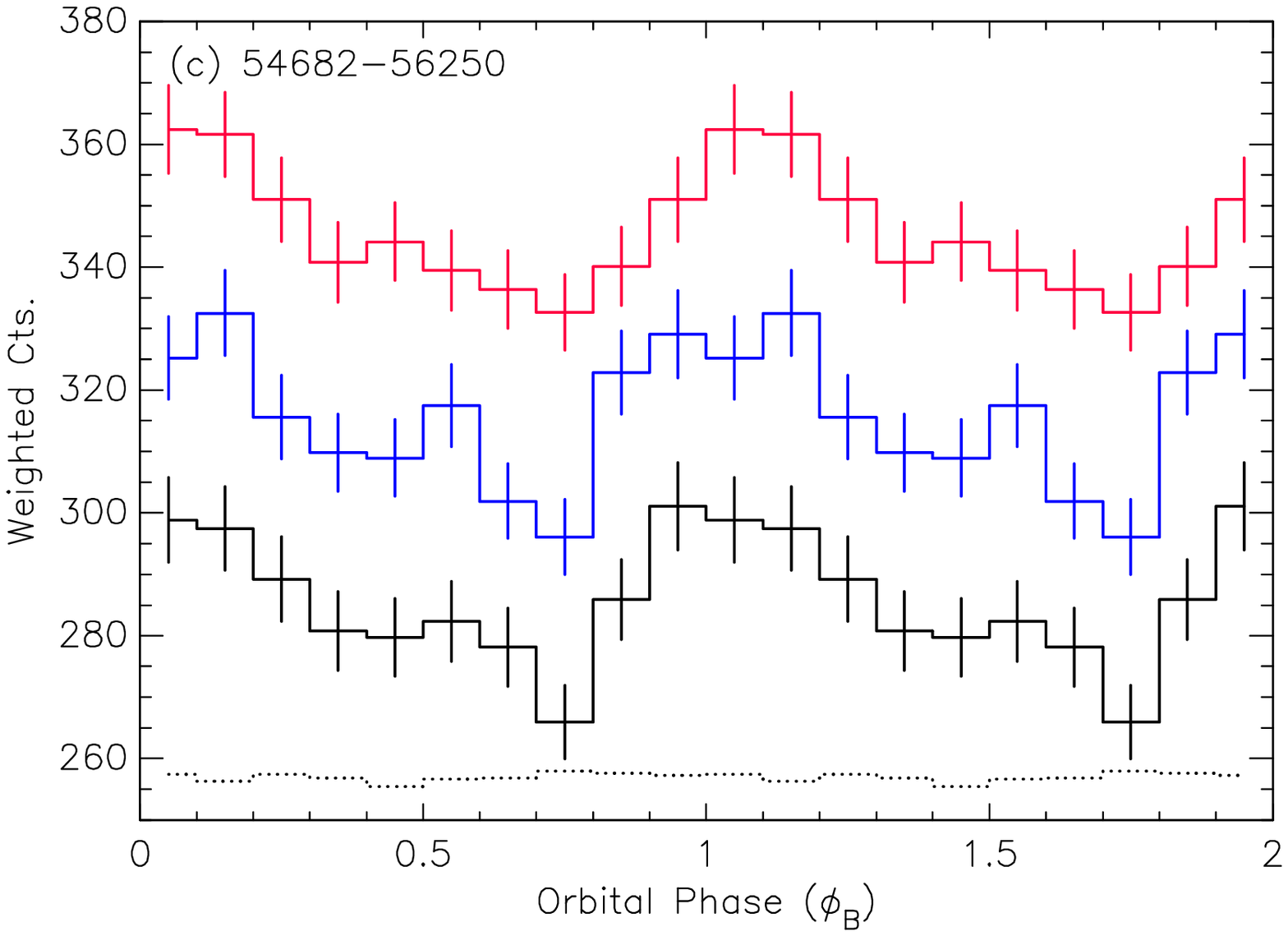} &
\includegraphics[width=3.3 in]{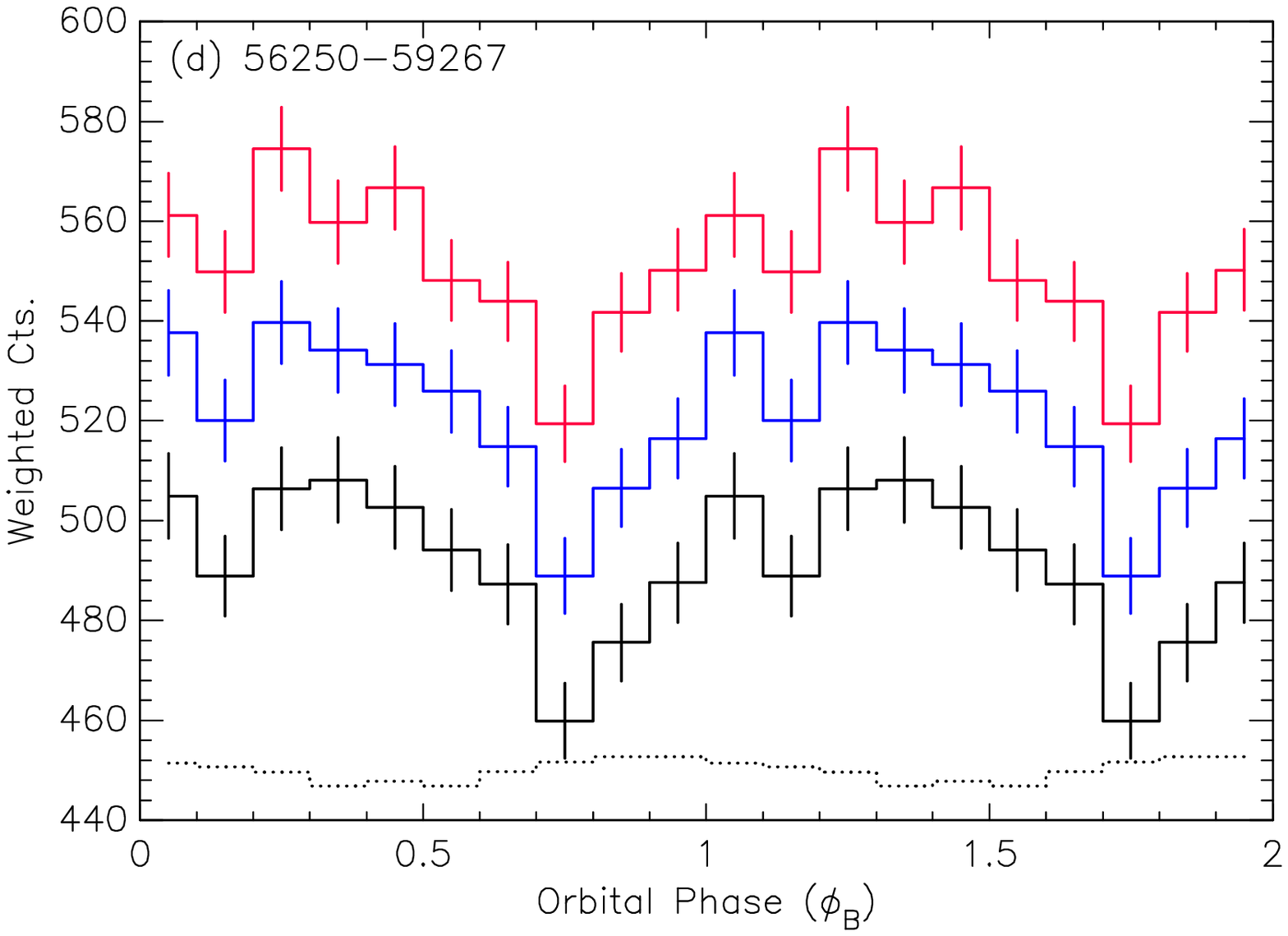} \\
\end{tabular}
\figcaption{Exposure-corrected 60\,MeV--1\,GeV orbital light curves
folded on the timing solutions in Table~\ref{ta:ta2}
(a, c, and d) and results for time-cumulative $H$ tests (b).
Black, blue, and red lines are for Solutions~1, 2, and 3, respectively, and
black dotted lines show exposure variation (scaled to show in the same panels with the light curves).
Time intervals for the light curves are shown in the upper left corner: full 12.5\,yr (a), pre-transition (c),
and post-transition (d) intervals.
The blue and red light curves are displaced vertically and zero is suppressed for better legibility.
$\phi_{\rm B}=0$ is the ascending node.
({\it b}) $H$-statistic values for the time interval $T_{\rm 0}$--$T_{\rm 1}$,
where $T_{\rm 0}$ is the start of the LAT mission and $T_{\rm 1}$ increases from
the start with a time step of $10^6$\,sec.
Red vertical lines denote the suggested range for the state transition, and
the horizontal arrows show the pre- and post-transition intervals used for
light curves in (c) and (d).
\label{fig:fig2}
}
\vspace{0mm}
\end{figure*}

\subsection{Timing Analysis}\label{sec:sec3_2}
For a timing analysis, we perform an unbinned likelihood analysis.
We process the data following the standard
procedure,\footnote{https://fermi.gsfc.nasa.gov/ssc/data/analysis/scitools/likeliho\\od\_tutorial.html}
compute a probability weight for each event using {\tt gtsrcprob} with the 4FGL spectral models,
and barycenter-correct the event arrival times with {\tt gtbary}.
Exposure is separately computed with {\tt gtexposure}.
We produce probability-weighted light curves by folding events within $R=3^\circ$ in the low-energy band (60\,MeV--1\,GeV)
because gamma-ray modulations have been seen better
at low energies \citep[e.g.,][]{ark18}; perhaps the modulating signal drops rapidly with energy.
We carry out a weighted $H$ test \citep[][]{k11} and
find significant modulation of the orbital light curves for all the
solutions in Table~\ref{ta:ta2} with $H$ statistics of 30, 30, and 31, for Solutions~1, 2 and 3, respectively,
corresponding to $p\approx5\times 10^{-6}$. Changing the RoI size or energy band slightly
does not alter the results significantly.
The three timing solutions give similar results (here and below), and so we report only the representative ones below.
Note that we also searched $>$1\,GeV data but did not find any significant modulation. Considering this
trial would make the detection significance lower by a factor of 2 (i.e., $p\approx2\times 10^{-6}$).

We further check to see if the low-energy modulation is induced by variation of exposure or variability
of a nearby source. We fold the exposure on the same timing solutions and find that orbital variation of the exposure
is only $<$1\% (black dotted lines in Figure~\ref{fig:fig2}). It shows a different shape (e.g., maximum at $\phi_{\rm B}$=0.75)
from the light curve and thus is unlikely to induce the observed modulation of J1227.
We also perform the same timing analysis for the nearby bright
source 4FGL~J1231.6$-$5116 ($\sim$2$^\circ$ from J1227) and find no significant modulation ($H\approx12$) in this source
at the orbital period of J1227. Besides, the light curve of 4FGL~J1231.6$-$5116 has
a different shape (e.g., phases of minimum and maximum) from that of J1227.
We also checked with other bright sources in the field (Fig.~\ref{fig:fig1x}) and verified that they
did not exhibit significant modulation either.

The low-energy (60\,MeV--1\,GeV) LAT light curves have a broad bump (Fig.~\ref{fig:fig2})
with a minimum at $\phi_{\rm B}\approx 0.75$ (IFC)
as opposed to the optical (Fig.~\ref{fig:fig1}) and X-ray light curves \citep[maximum at IFC;][]{dpbp+20}.
This is the opposite of the result from a previous gamma-ray study
(0.3--300\,GeV; XW15) in which
the gamma-ray maximum was suggested to be at IFC.
We inspect 0.3--300\,GeV data
but do not find any significant modulation in that band (e.g., $H\approx4$).

We check to see if there is energy dependence in the modulation. We split the
0.6\,MeV--1\,GeV band into two energy bands (60\,MeV--450\,MeV and 450\,MeV--1\,GeV) with a similar $H$ value
($H\approx$15 in each band). The shape of the light curve
seems not to change with energy; they look almost the same as that in Figure~\ref{fig:fig1} (a).
However, the modulation fractions in the two sub-bands differ:
$\approx$4\% and $\approx$10\% in the 60\,MeV--450\,MeV and 450\,MeV--1\,GeV bands, respectively.
This implies that the modulating emission becomes weaker with energy less rapidly
than the orbitally-constant pulsed emission does; the latter
exponentially cuts off at high energies.

We perform a time-cumulative $H$ test with the 60\,MeV--1\,GeV data
by gradually increasing the time interval for the test from the start
to the end of the LAT data,
and find that the significance for modulation increases with time nearly monotonically except for
some relatively short intervals. The result seems to suggest that
the modulation is quite strong even ``before'' the transition ($<$MJD~56250) with an $H$ value of
$\sim$20. In order to confirm this we split the observation into two
intervals: a pre-transition and a post-transition period
(before and after MJD~56250; see blue arrows in Fig.~\ref{fig:fig2} b),
and compute $H$-statistic values for orbital modulation in each of them
in the 60\,MeV--1\,GeV band.
We find $H\approx$19 ($p\approx$5$\times10^{-4}$)
and $H\approx$17 ($p\approx$$10^{-3}$)
in the pre- and post-transition data, respectively (Fig.~\ref{fig:fig2} c and d).
The modulation fraction $(F_{\rm max} - F_{\rm min})/(F_{\rm max} + F_{\rm min})$ is
approximately 5\% for all the pre-, post-transition, and time-integrated light curves.
This possibly suggests that the gamma-ray modulation persists in both LMXB and MSP states of the tMSP J1227.

\section{Discussion}
\label{sec:sec4}
Because there is no long-term timing solution for the pulsar in the tMSP system
J1227, we extrapolated radio timing solutions obtained over short time intervals
(less than 1000\,days) for the 15-year optical data and verified that the orbital solutions
phase-align well the optical data.
Hence, we used the extrapolated solutions for analyses of {\it Fermi} LAT data,
and discovered significant ($p\approx 5\times10^{-6}$) orbital modulation in the
low-energy (60\,MeV--1\,GeV) gamma-ray emission of J1227.
The light curves of the source have a minimum near
IFC ($\phi_{\rm B}=0.75$), which is different from a previous
suggestion that the gamma-ray modulation of J1227 has a maximum at that phase.
We further found that the source seemed to exhibit similar modulation
in the pre- and post-transition periods.
The time-resolved light curves (Fig.~\ref{fig:fig2} c and d),
having a minimum at IFC, are similar to the time-integrated one.
No significant gamma-ray modulation was detected at higher energies ($>$300\,MeV).

XW15 claimed 3-$\sigma$ gamma-ray modulation ($H\approx 11$)
in the $>$300\,MeV band assuming an orbital frequency of $4.018\times 10^{-5}\rm \ Hz$
in a post-transition period, which we were not able to reproduce.
Although the discrepancy may be due to differences in the datasets (``Pass 8'' vs ``Pass 7''),
an analysis carried out with ``Pass 7'' data did not reproduce
the $>$300\,MeV modulation either \citep[][]{jrrc+15}.
Because the orbital frequency assumed by XW15 is not consistent with the
radio solutions (Table~\ref{ta:ta2}) and one we found by fitting optical light curves (\S~\ref{sec:sec2_2}),
and the detection significance is not high (would be lower considering some trial factors),
we speculate that the $>$300\,MeV modulation might be spurious.

Our investigation of the radio timing solutions suggests that higher-order
derivatives (e.g., higher than the second time derivatives) of the orbital
frequency seem to be small as in PSR~J2039$-$5617
\citep[$\Delta P_{\rm B}/P_{\rm B}\le 10^{-6}$;][]{cnva+21}.
Otherwise, it is very difficult to maintain the
phase alignment of the optical data over the 15\,yr period (e.g., Fig.~\ref{fig:fig1}).
Our inspection of the timing solutions does not provide a definite test for them,
and the current solutions may be slightly inaccurate.
Small inaccuracy \citep[e.g., $\sim$10\,s phase offsets for PSR~J2039$-$5617;][]{cnva+21}
would not change our conclusions significantly since
the gamma-ray modulation is detected regardless of the timing solutions we tested
and so seems to be quite robust.
However, phases of the gamma rays might have been mixed, which makes the light curve broader;
the intrinsic gamma-ray light curve of J1227 may be somewhat narrower. The phase mixing is particularly
a concern for a phase-resolved spectroscopy; a phase-resolved study awaits a better timing solution.

In time-cumulative $H$ tests (\S~\ref{sec:sec3_2}) for the low-energy modulation, we find that the
significance for it (i.e., $H$ value) seems to increase more rapidly
between MJDs~55400 and 56250 than at other times (Fig.~\ref{fig:fig2}).
XW15 noted that long-term ($>$30\,day) variability in J1227 was strong in MJD~55400--55800
and the source was brighter in MJD~55800--56250 than before.
Although it is unclear whether the stronger modulation and long-term variability
in the time interval are related to each other, it will be intriguing to see if a physical
scenario could explain a simultaneous change
of orbital modulation and long-term variability in the LMXB state.
Then, the large drop in cumulative $H$ right after the transition (Fig.~\ref{fig:fig2})
perhaps indicates that the source might have relaxed back to a stationary state
with the transition.

While gamma-ray emission in the MSP state is dominated by magnetospheric
(pulsed) emission which is orbitally constant, the detection of orbitally modulating
gamma rays in the MSP state of J1227 suggests that there are other emission components.
The orbital modulation in the gamma-ray emission of J1227 and other pulsar
binaries \citep[e.g.,][]{wtch+12,arjk+17,ark18,ntsl+18,ark20,cnva+21} can be
explained with an IBS scenario. In the scenario \citep[e.g.,][]{whvb+17,kra19,mwvh+20},
the alignment of optical and X-ray orbital phasing of J1227 (i.e., maximum at IFC)
implies that the IBS wraps around the pulsar.
If the modulating gamma rays are produced primarily via ICS of the stellar seed photons by electrons
in the IBS or the pre-shock wind, the gamma-ray minimum is expected at IFC due to
unfavorable viewing geometry. In the ICS-by-IBS case, however,
the ICS emission can be enhanced by Doppler beaming if the IBS flow has large bulk speed
and observer's line of sight is aligned with the flow (i.e., at IFC).
Then gamma-ray modulation would have a `maximum'
at IFC. The {\it Fermi}-LAT light curves of J1227, with a minimum near IFC,
suggest that ICS by the pre-shock wind is the main contributor to the gamma-ray modulation
or the bulk Doppler factor in the IBS flow is small.

More intriguingly, we find that J1227 seems to exhibit similar orbital modulation
even in its `LMXB' state (i.e., with accretion).
This is very puzzling because previous models for gamma-ray
emission in LMXB states seem not to predict
such modulation whether the emission mechanism is ICS of disk photons by
the pre-shock wind \citep[e.g.,][]{tllk+14} or synchrotron-self-Compton
at the magnetosphere-disk boundary \citep[e.g.,][]{ptl14}; these models
may explain the stronger (i.e., compared to the MSP state) gamma-ray emission
and its long-term variability in the LMXB state \citep[e.g., $>$30-day scale;][XW15]{jrrc+15}
but do not provide varying conditions
in circular orbits. Because the phasings (i.e., minimum at IFC) and modulation
fractions of the gamma-ray light curves in the LMXB and MSP states are similar,
we speculate that ICS of the stellar photons by the pre-shock pulsar wind
or IBS particles is an important contributor to the production of modulating
gamma-ray emission in the LMXB state.

Note, however, that the pre- and post-transition light curves (Figs.~\ref{fig:fig2} c and d)
appear qualitatively different (e.g., narrower in the LMXB state), indicating
that there may be some physical differences between the two states of the source.
Perhaps, the disk may play some role in the production of gamma-ray modulation.
It is difficult to investigate this further because of poor quality of the light-curve measurements.
Moreover, the significance of the modulation in each of
the LMXB and the MSP states is yet low. Further confirmations are needed, and multiwavelength data and
theoretical models can help to understand the physics behind the state change better.

It will be interesting to see if the orbital modulation of J1227 is `pulse-phase' dependent
as in PSR~J2339$-$0533 \citep[][]{ark20}. Orbital-phase-resolved spectroscopy can
help to measure the ICS spectrum and
model the multi-band modulation with
an IBS \citep[][]{tllk+14,ar17} and/or a propeller scenario \citep[][]{ptl14}.
These demand a more accurate timing solution that is valid through the LAT operation.
While pulsar binaries are hard to time, optical data may help radio and LAT
timing studies by providing a guide to orbital solutions.

\bigskip
\bigskip

\acknowledgments
The \textit{Fermi} LAT Collaboration acknowledges generous ongoing support from a number
of agencies and institutesthat have supported both the development and the operation of
the LAT as well as scientific data analysis. These include the National Aeronautics and
Space Administration and the Department of Energy in the United States,
the Commissariat \`a l'Energie Atomique and the Centre National de la Recherche
Scientifique / Institut National de Physique Nucl\'eaire et de Physique des Particules
in France, the Agenzia Spaziale Italiana and the Istituto Nazionale di Fisica Nucleare
in Italy, the Ministry of Education, Culture, Sports, Science and Technology (MEXT),
High Energy Accelerator Research Organization (KEK) and Japan Aerospace Exploration
Agency (JAXA) in Japan, and the K.~A.~Wallenberg Foundation, the Swedish Research
Council and the Swedish National Space Board in Sweden.

Additional support for science analysis during the operations phase is gratefully
acknowledged from the Istituto Nazionale di Astrofisica in Italy and the
Centre National d'\'Etudes Spatiales in France.
This work performed in part under DOE Contract DE-AC02-76SF00515.

This research was supported by Basic Science Research Program through the National
Research Foundation of Korea (NRF) funded by the Ministry of Science, ICT \& Future
Planning (NRF-2017R1C1B2004566).

\vspace{5mm}
\facilities{{\it Swift}, {\it XMM-Newton}, {\it Fermi}}
\software{{\it Fermi} ST (v1.2.23),  HEASoft \citep[v6.27;][]{heasarc2014}, XMM-SAS \citep[v20190531;][]{xmmsas17}}

\bibliographystyle{apj}
\bibliography{ms}
\end{document}